# An accurate algorithm for calibration-free wavelength modulation spectroscopy based on even-order harmonics

Yihong Wang[a], Bin Zhou[b], Bubin Wang[c], Rong Zhao[d]

School of Energy and Environment, Southeast University, Nanjing 210096, China

[a]wyh@seu.edu.cn, [b]zhoubinde@seu.edu.cn, [c]wbb@seu.edu.cn, [d]zhaorong@seu.edu.cn

**Abstract.** This paper proposes an accurate algorithm to implement calibration-free wavelength modulation spectroscopy based on even-order harmonics. The proposed algorithm, analytically deduced from a much more accurate Voigt function model, enabling not only speedy measurement down to milliseconds, but also a general suitability for various degrees of line-shape broadening. The proposed method is validated by condition-controlled experiment, indicating calculation of gas temperature with the relative error less than 2.4% in the experiments.

**Introduction**

Wavelength modulation spectroscopy (WMS), which is a modality of the tunable diode laser absorption spectroscopy (TDLAS), based on harmonic detection has been widely employed for gas properties measurement, given its advantages of contactless, rapid, high sensitivity and accuracy[1, 2]. To achieve calibration-free WMS (CF-WMS), attempts have been made in terms of 1f-normalization [3-6], Fourier-based multi-parameter harmonic fitting [7], absorbance lineshape recovery [8-10] and quasi-simultaneous direct absorption spectroscopy (DAS) detection [10]. Among all the above-mentioned techniques, the calibration-free $nf/1f$ method [4-6], noted as CF-$nf/1f$, that belongs to the 1f-normalization strategy, is most widely applied due to its ease of operation, free from complex analytic models, and general suitability for various measuring conditions, e.g. elevated temperature/pressure [11]. When implementing the CF-$nf/1f$ method, the time-dependent measured or simulated incident laser intensity, i.e. $I_0^S(t)$ in Fig.1, and time-dependent wavelength modulation frequency response (WMFR), i.e. $v(t)$, need to be measured in advance. According to the Beer's law, the simulated transmitted laser intensity, i.e. $I_t^S(t)$, can be subsequently obtained for given gas parameters, e.g. concentration $X$, temperature $T$, transition linecenter frequency $v_0$, integrated absorbance area $A$, Doppler width $\Delta v_D$, collisional width $\Delta v_C$, etc. With the measured transmitted laser intensity, i.e. $I_t^M(t)$, and $I_t^S(t)$ in hand, the gas properties can be inferred when the sum-of-squared errors (SSE) or the amplitude difference between the measured and simulated WMS signals reaches the minimum. A brief flow chart describing the CF-$nf/1f$ method is shown in Fig. 1.

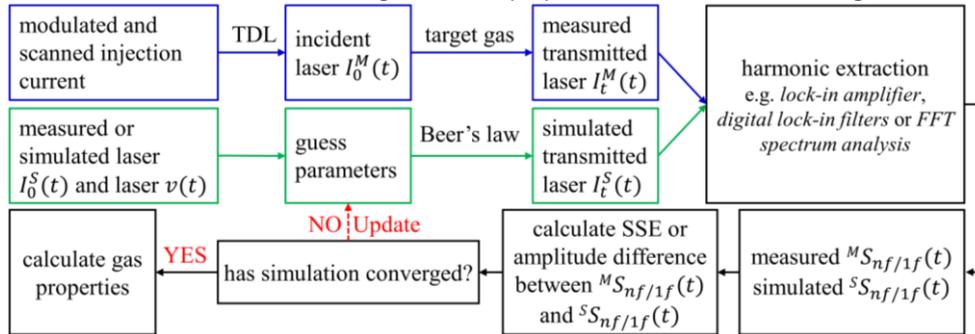

Fig. 1. A brief flow chart describing the CF-nf/1f method.

However, accurate retrieval of the simulated $nf/1f$, noted as $^Snf/1f$ in Fig. 1, is still challenging since the harmonic signals are influenced by many factors, such as nonlinear variations in laser characteristics and laser temperature variations [11, 12]. In order to solve the problems mentioned above, we recently developed a novel and rapid algorithm, denoted as *Algorithm I* in this paper, for calibration-free wavelength modulation spectroscopy based on even-order harmonics[13]. However, *Algorithm I* is built upon the approximate description of the Voigt function proposed by Liu et al[14]. This approximation has a given validity. As long as the methodology is based on this, the accuracy of the retrieved parameters cannot be better than the accuracy of this approximation. Therefore, an updated *Algorithm II* built upon much more accurate Voigt approximation is necessary.

**Methodology**

**Theoretical Background.** The detailed theory of *Algorithm I* can be found in reference[13]. In this subsection, only some main conclusions are listed here. It is worth noting that the mathematical symbols in this paper are consistent with reference[13]. The spectroscopic absorbance $\alpha(v)$ can be written as:

$$\alpha(v) = -\ln(I_t / I_0) = A\varphi(v), \tag{1}$$

where $v$ is the instantaneous laser frequency, $I_t$ and $I_0$ are the transmitted and incident laser intensities, $A$ is the integral absorption area, $\varphi(v)$ is the line-shape function. In *Algorithm I*, the spectroscopic absorbance was approximated by:

$$\varphi(v) = \frac{2}{\pi\lambda}[c_L L(\Delta, m) + c_G \sqrt{\pi \ln 2} G(\Delta, m)], \tag{2}$$

where $\lambda$ represents the full width at half maximum (FWHM) of the absorption line, $m=2a/\lambda$ represents the modulation index, $a$ is the modulation depth, $\Delta = 2(v - v_0)/\lambda$ is introduced to describe the relative offset of the laser center frequency $v_0$, $L(\Delta, m)$ and $G(\Delta, m)$ are the time-dependent peak normalized Lorentzian and Gaussian linear functions respectively, $c_L$ and $c_G$ are the weights of the Lorentzian and Gaussian broadening coefficients. In this work, a much more accurate model for the line-shape function described by Voigt function is used for harmonic calculation[15]:

$$\varphi(v) = \frac{2}{\lambda_G}\sqrt{\frac{\ln 2}{\pi}} K(x, y) \tag{3}$$

where $\lambda_G$ is the Gaussian broadening, $K(x,y)$ is the Voigt function defined as:

$$K(x, y) = \frac{y}{\pi}\int_{-\infty}^{+\infty} \frac{e^{-t^2}}{(x-t)^2 + y^2} dt \tag{4}$$

where $x$ and $y$ are

$$x = \sqrt{\ln 2}\,\frac{v - v_0}{\lambda_G} \tag{5}$$

$$y = \sqrt{\ln 2}\,\frac{\lambda_L}{\lambda_G} \tag{6}$$

where $\lambda_L$ is the Lorentzian broadening. In the WMS technique, parameter $x$ can be reformulated as:

$$x = \sqrt{\ln 2}\,\frac{a\cos(\omega t)}{\lambda_G} = 2\sqrt{\ln 2}\,\frac{\lambda}{\lambda_G} m\cos(\omega t) \tag{7}$$

where $\omega$ is the angular frequency of sinusoidal modulation, $t$ means time. In order to simplify the analysis, a line-shape parameter $d$ is introduced as:

$$d = \frac{\lambda_L - \lambda_G}{\lambda_L + \lambda_G} \tag{7}$$

Considering that $\lambda$ is determined by $\lambda_L$ and $\lambda_G$[16], both parameter $y$ and parameter $\lambda/\lambda_G$ can be represented by parameter $d$:

$$y = \sqrt{\ln 2}\,\frac{1+d}{1-d} \tag{9}$$

$$\frac{\lambda}{\lambda_G} = f(d) \tag{10}$$

where $f(.)$ is an anonymous function. Thus the exact mathematical model of the spectroscopic absorbance $\alpha(v)$ represented by parameters $m$ and $d$ can be expressed as:

$$\alpha(v) = \frac{A}{a} F(m,d) \tag{11}$$

where $F(m,d)$ is defined as:

$$F(m,d) = mf(d)\sqrt{\frac{\ln 2}{\pi}} K[2\sqrt{\ln 2}\, mf(d)\cos(\omega t), \sqrt{\ln 2}\,\frac{1+d}{1-d}] \tag{12}$$

Obviously, the harmonics of $\alpha(v)$ and $F(m,d)$ are proportional. The amplitude of the nth harmonic of $F(m,d)$ can be expressed as:

$$H_n = \frac{\omega}{\pi} \int_{-\frac{\omega}{\pi}}^{\frac{\omega}{\pi}} F(m,d)\cos(n\omega t)dt \tag{13}$$

Unfortunately, since there is no analytically exact expression to describe the Voigt function, the nth harmonic amplitude $H_n$ also have no analytically exact expression.

We note that high precision value of $H_n$ for given $m$ and $d$ can be calculated by numerical method. These amplitude databases $H_n(m,d)$ for the 2nd-, 4th-, 6th- and 8th- harmonic are show in Fig. 2(a), 2(a), 2(c) and 2(d), respectively. The resolution of parameters $m$ and $d$ both are 0.01 for these established databases. Therefore, the accurate estimation of $H_n$ under arbitrary $m$, $d$ can be derived base on spline interpolation method. Numerical results show that the relative deviation of the estimated $H_n$ is less than $2.37 \times 10^{-6}$ when $-1 < d \leq 1$ and $2.5 \leq m \leq 4.5$. It is worth noting that the ranges of $m$ and $d$ in this paper are selected based on the optimal parameters of *Algorithm I*. For other parameter ranges or higher precision requirements, these databases can be established in the same way. In addition, the raw data with double precision in Fig. 2 can be found at the MATLAB Central website: https://www.mathworks.com/matlabcentral/fileexchange/101554-database-for-wavelength-modulation-spectroscopy-algorithm.

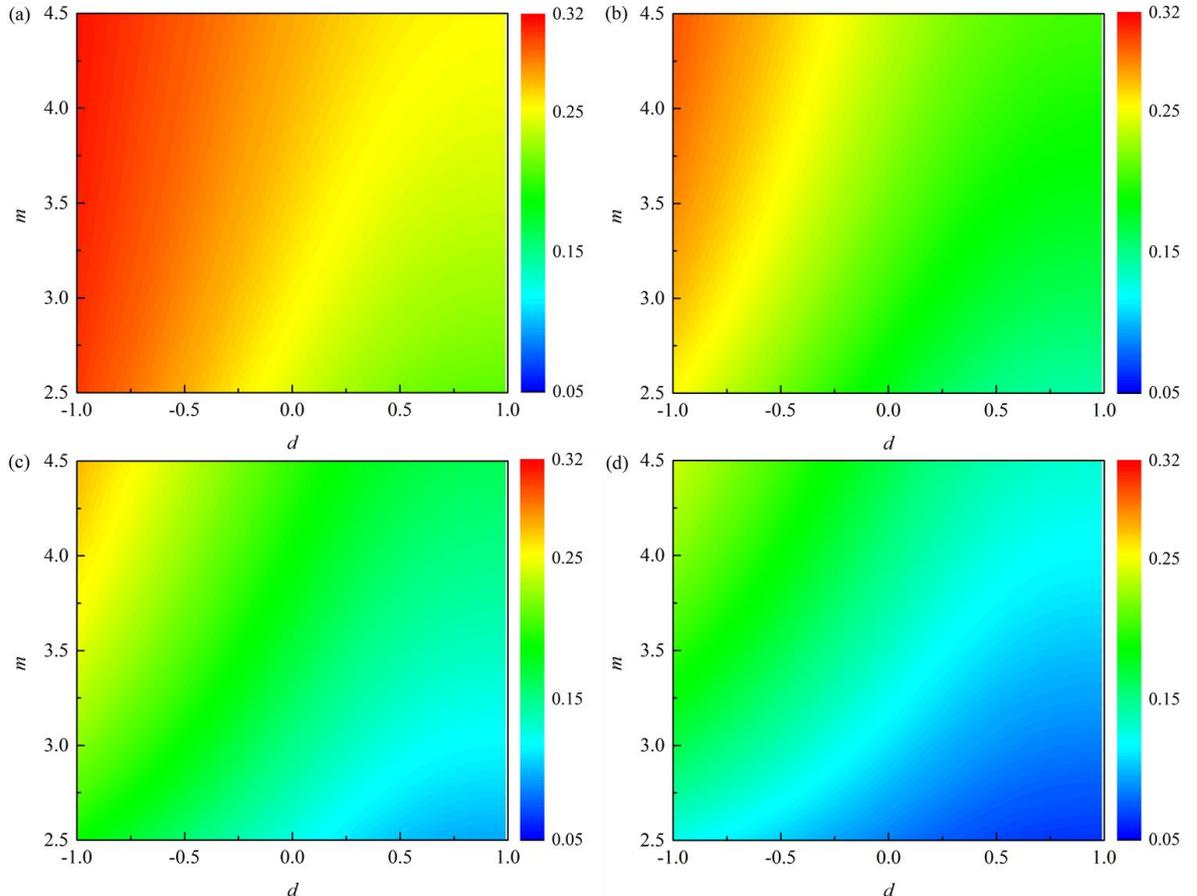

Fig. 2. Amplitude databases $H_n(m,d)$ for the (a) 2nd-, (b) 4th-, (c) 6th- and (d) 8th- harmonic, respectively.

**Fundamentals of the proposed *Algorithm II*.** Fig. 2 show that the amplitude of $H_n$ decrease as the even harmonic order $n$ increases. Consequently, the first several nonzero even harmonics should be adopted to achieve higher signal-to-noise ratios (SNRs) of the WMS measurement. In *Algorithm II*, the parameters $m$ and $d$ can be calculated by solving the following minimization problem:

$$(m,d) = \arg\min_{m,d}\{[\frac{h_4}{h_2} - \frac{H_4(m,d)}{H_2(m,d)}]^2 + p_1[\frac{h_6}{h_2} - \frac{H_6(m,d)}{H_2(m,d)}]^2 + ... + p_k[\frac{h_{2k+4}}{h_2} - \frac{H_{2k+4}(m,d)}{H_2(m,d)}]^2\} \quad (14)$$

where $h_2$, $h_4$, $h_6$, …, are the measured harmonics amplitudes, $p_1$, $p_2$, …,$p_k$ are the corresponding weight coefficient. In this work, only the first three nonzero even harmonics are adopted and the corresponding weight coefficient are both set to 1. Therefore, Eq. 14 can be reformulated as:

$$(m,d) = \arg\min_{m,d}\{[\frac{h_4}{h_2} - \frac{H_4(m,d)}{H_2(m,d)}]^2 + [\frac{h_6}{h_2} - \frac{H_6(m,d)}{H_2(m,d)}]^2\} \quad (15)$$

With the parameters $m$ and $d$ in hand, the FWHM of the Voigt line-shape can be calculated by

$$\lambda = 2a/m \quad (16)$$

and thus the integrated absorbance area $A$ can be calculated by

$$A = \frac{ah_2}{H_2(m,d)} \quad (17)$$

The integrated absorbance areas obtained from two individual transitions can be used to infer the gas temperature using two-line thermometry [17, 18]. With the temperature in hand, the gas concentration can be calculated from one of the integrated absorbance areas.

**Experimental Verification**

For a demonstrative purpose, we process the experimental data that has been published in our recent work[13] with the *Algorithm II* introduced. The schematic diagram of the experimental system is shown as Fig. 3.

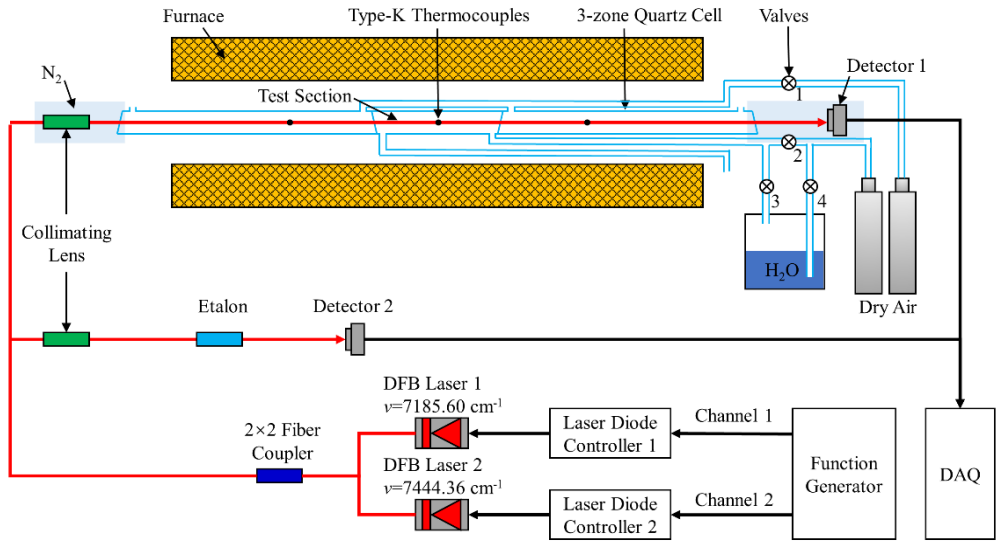

Fig. 3. Schematic diagram of the experimental system.

Since the details of the experiments setup can be find in reference[13], we directly give the comparison of temperature measurement results with different algorithms here. The calculated temperature values are compared with those measured by the thermocouple in the range of 773 - 1273 K with an interval of 100 K. As shown in Fig. 4 (a), the temperature calculated in all the two algorithms agree well with that measured by the thermocouple. Fig. 4 (b) show that the maximum relative errors are 2.59% and 2.40% for *Algorithm I* and *Algorithm II*, respectively. Experimental results show the effectiveness of the proposed algorithm.

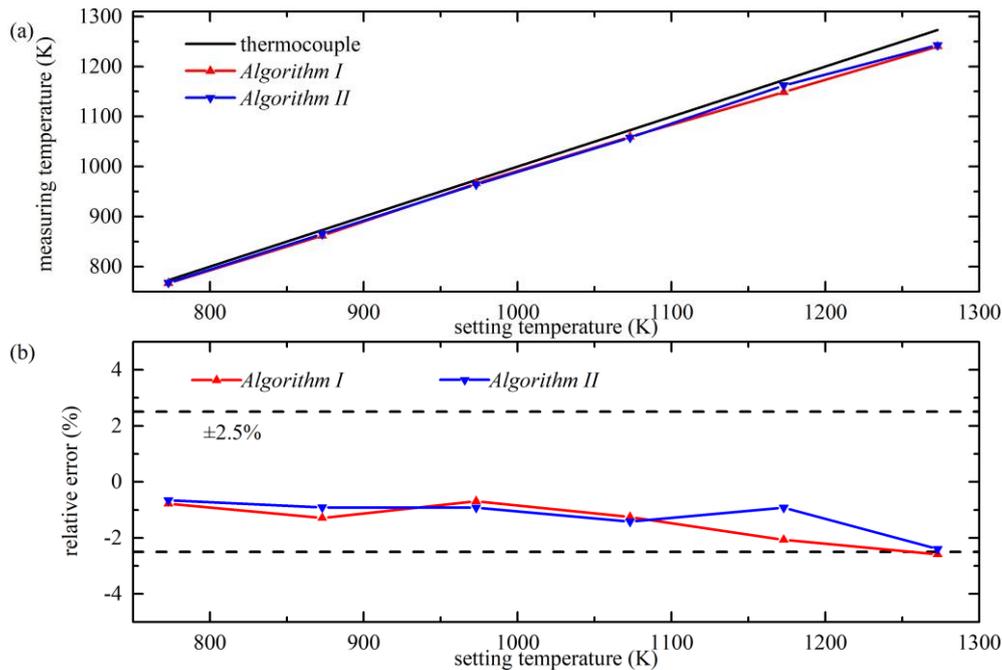

Fig. 4. Comparison of the temperature measured and the thermocouple. (a) and (b) show the absolute measured temperature and the relative error, respectively.

**Summary**

To achieve calibration-free WMS, an accurate algorithm is proposed for accurate retrieval of gas properties based on the even-order harmonics. This algorithm, which is built upon much more accurate Voigt approximation, is an updated version of *Algorithm I* proposed in our recent work. The proposed *Algorithm II* is validated by condition-controlled experiment, indicating calculation of gas temperature with the relative error less than 2.4% in the experiments.

**Funding.** National Key Research and Development Program of China (2017YFB0603204)

**References**

[1] C. Liu and L. Xu, "Laser absorption spectroscopy for combustion diagnosis in reactive flows: A review," Applied Spectroscopy Reviews **54**, 1-44 (2019).

[2] Y. Wang, B. Zhou, and C. Liu, "Sensitivity and Accuracy Enhanced Wavelength Modulation Spectroscopy Based on PSD Analysis," IEEE Photonics Technology Letters, doi: 10.1109/LPT.2021.3128448.

[3] G. B. Rieker, J. B. Jeffries, and R. K. Hanson, "Calibration-free wavelength-modulation spectroscopy for measurements of gas temperature and concentration in harsh environments," Applied optics **48**, 5546-5560 (2009).

[4] K. Sun, X. Chao, R. Sur, C. S. Goldenstein, J. B. Jeffries, and R. K. Hanson, "Analysis of calibration-free wavelength-scanned wavelength modulation spectroscopy for practical gas sensing using tunable diode lasers," Measurement Science and Technology **24**, 125203 (2013).

[5] C. S. Goldenstein, C. L. Strand, I. A. Schultz, K. Sun, J. B. Jeffries, and R. K. Hanson, "Fitting of calibration-free scanned-wavelength-modulation spectroscopy spectra for determination of gas properties and absorption lineshapes," Applied optics **53**, 356-367 (2014).

[6] J. Vanderover, W. Wang, and M. A. Oehlschlaeger, "A carbon monoxide and thermometry sensor based on mid-IR quantum-cascade laser wavelength-modulation absorption spectroscopy," Applied Physics B **103**, 959-966 (2011).

[7] Y. Zakrevskyy, T. Ritschel, C. Dosche, and H. G. Löhmannsröben, "Quantitative calibration- and reference-free wavelength modulation spectroscopy," Infrared Physics & Technology **55**, 183-190 (2012).

[8] K. Duffin, A. J. McGettrick, W. Johnstone, G. Stewart, and D. G. Moodie, "Tunable Diode-Laser Spectroscopy With Wavelength Modulation: A Calibration-Free Approach to the Recovery of Absolute Gas Absorption Line Shapes," J. Lightwave Technol. **25**, 3114-3125 (2007).

[9] G. Stewart, W. Johnstone, J. R. P. Bain, K. Ruxton, and K. Duffin, "Recovery of Absolute Gas Absorption Line Shapes Using Tunable Diode Laser Spectroscopy With Wavelength Modulation—Part I: Theoretical Analysis," J. Lightwave Technol. **29**, 811-821 (2011).

[10] Z. Peng, Y. Du, and Y. Ding, "Highly Sensitive, Calibration-Free WM-DAS Method for Recovering Absorbance—Part I: Theoretical Analysis," **20**, 681 (2020).

[11] Y. Du, Z. Peng, and Y. Ding, "A high-accurate and universal method to characterize the relative wavelength response (RWR) in wavelength modulation spectroscopy (WMS)," Optics Express **28**, 3482-3494 (2020).

[12] Z. Wang, P. Fu, and X. Chao, "Laser Absorption Sensing Systems: Challenges, Modeling, and Design Optimization," **9**, 2723 (2019).

[13] Y. Wang, B. Zhou, and C. Liu, "Calibration-free wavelength modulation spectroscopy based on even-order harmonics," Optics Express **29**, 26618-26633 (2021).

[14] Y. Liu, J. Lin, G. Huang, Y. Guo, and C. Duan, "Simple empirical analytical approximation to the Voigt profile," JOSA B **18**, 666-672 (2001).

[15] N. Mohankumar and S. Sen, "On the very accurate evaluation of the Voigt functions," Journal of Quantitative Spectroscopy & Radiative Transfer **224**, 192-196 (2019).

[16] J. J. Olivero and R. L. Longbothum, "Empirical fits to Voigt line width : brief review," Journal of Quantitative Spectroscopy & Radiative Transfer **17**, 233-236 (1977).

[17] C. Liu, L. Xu, J. Chen, Z. Cao, Y. Lin, and W. Cai, "Development of a fan-beam TDLAS-based tomographic sensor for rapid imaging of temperature and gas concentration," Optics Express **23**, 22494-22511 (2015).

[18] C. Liu, L. Xu, Z. Cao, and H. McCann, "Reconstruction of Axisymmetric Temperature and Gas Concentration Distributions by Combining Fan-Beam TDLAS With Onion-Peeling Deconvolution," IEEE Transactions on Instrumentation and Measurement **63**, 3067-3075 (2014).